      [1999/12/01 v1.4c Il Nuovo Cimento]
\documentclass{cimento}

\title{Tevatron Legacy}
\author{Y.~Peters\from{ins:x}\ETC\thanks{On behalf of the D0 and CDF
    collaborations. Fermilab preprint number: FERMILAB-CONF-11-662-PPD}}
\instlist{\inst{ins:x} University of Manchester\-Oxford Road,
  Manchester, UK}
\PACSes{\PACSit{14.65.Ha}{Top Quarks}}
\begin{document}

\maketitle

\begin{abstract}
Several major milestones and discoveries were attained during the lifetime of the Tevatron proton-antiproton collider at
Fermilab, from 1987 to 2011. One of the most important was the discovery of the top quark
in 1995, followed by an  intense program to study that  particle
in greater detail. In this article, I give an overview of the history of the top
quark, its current status as well as the still
to be completed legacy measurements at the Tevatorn.
\end{abstract}

\section{Introduction}
The heaviest fundamental particle ever known, the top quark,
was discovered at the Tevatron proton antiproton ($p\bar{p}$) collider at
Fermilab in 1995. Since then, intensive studies continue to be  performed at the
two general purpose detectors at the Tevatron, CDF and D0, to gain a better
understanding of the production mechanisms and properties of that 
particle. With the start of operations of the Large Hadron Collider (LHC)
at CERN in 2010, a top quark factory has opened, where top-antitop quark pairs ($t\bar{t}$)
are currently produced at cross sections a factor of about twenty higher than at the Tevatron,
raising the question about  the legacy of the Tevatron, i. e., the
heritage the Tevatron leaves to the physics community. 
In this article, I offer a brief review of the history of the Tevatron
and of the top
quark, an overview of the current status of the understanding of the top quark, as
well as a perspective of what measurements are still important to complete
in top-quark physics at the Tevatron in view of the onset  of the LHC-era.

\section{The Past}
From its start, until the turn-on of the LHC, the
Tevatron  $p\bar{p}$ collider has been the highest-energy frontier of
the world. 
During the lifetime of the  Tevatron, which was commissioned in 1983, and switched off on September 30th, 2011, several important
milestones have been attained, as, for example, the observation of the top quark~\cite{cdftopdiscovery, d0topdiscovery} and
the detailed study of its properties,
the observation of $B_S$ oscillations, the precise measurement of the
mass of the $W$ boson and the extensive hunt for the Higgs boson, that
led to
stringent limits over a wide range of Higgs-boson mass values. All of
these discoveries and studies have enhanced 
 our understanding of nature. In this section, I briefly review the history of
the Tevatron and that of the top quark.

\subsection{A brief history of the Tevatron}
With the goal to double the energy of the existing accelerator at
Fermilab, and thereby explore new regions of phase space, the planning
to build the Tevatron
$p\bar{p}$ collider  using superconducting magnets was initiated in
the late 1970s~\cite{TevHistory1, TevHistory2}. In summer 1983, the
Tevatron was completed and commissioned as fixed-target accelerator,
while the completion of the antiproton source and therefore the
initialization of first $p\bar{p}$ collisions in the Tevatron took
until 1985. In October 1985, the partially finished CDF detector
observed its first collisions. At this time, the D0 detector was still
under construction. In 1987, Run0 of the Tevatron started, at a
center of mass (CM) energy of 1.8~TeV. During Run0, only the
CDF detector was in operation, recording about 5~pb$^{-1}$ of data. To increase luminosity, the Tevatron and accelerator
complex were upgraded through a new  LINAC, introduction of transverse stochastic cooling in
the antiproton source, and installation of electrostatic separators in the
Tevatron. These improvements were associated with RunI of the
Tevatron, that
lasted from 1992 to 1996, at a CM energy of 1.8~TeV, but now with both
CDF and D0 detectors, in operation. During this time, namely in 1995,
the top quark was discovered. In total, CDF and D0 each recorded about
120~pb$^{-1}$ of data during RunI. After the end of RunI, the complex was
upgraded again, now with the introduction of the Main Injector and
Recycler. The invention and further development of electron cooling in the
Recycler enabled the
storage of large numbers of antiprotons, resulting in luminosities
well beyond initial expectation. At the same
time, both CDF and D0 detectors underwent upgrades.  With an
instantaneous 
luminosity goal of  a factor of about five greater than  that of RunI,
as well as a higher CM energy
of 1.96~TeV, RunII of the Tevatron started in March 2001, lasting
until September 2011. Both detectors collected about 10~fb$^{-1}$ of
data, that are currently being analysed.  

All in all, for many years the Tevatron was not only the highest-energy particle
collider in the world, but also a great symbol of  technological
innovation and breakthroughs that included the construction of superconducting magnets and
the introduction of stochastic cooling, and development of major
particle-detection and triggering schemes, all of which form part of the
legacy of the Tevatron.

\subsection{A brief history of the top quark}
One of the major achievements of the Tevatron was the discovery of the
top quark~\cite{TopHistory}. The previous discoveries of the upsilon and
thereby the $b$-quark at
Fermilab in 1977, together with the $\tau$-lepton at SLAC in 1976,
initialized the extension of quark generations in the Standard Model
(SM) of particle physics to three families. As  the  pattern
for the SM suggested  that a quark  of the third generation was
missing from an expected doublet, in particular, the up-type partner of
the $b$-quark, a race to find  this predicted particle, named the top quark,
was started. Based on the ratio of masses of the partners in the two known
families, it was widely believed that the top quark would have a mass
about a factor of three to four that of the $b$-quark. Several searches were
performed based on the expectation of a heavy new particle in a
$t\bar{t}$ bound state, such as the search at the PETRA experiment at the
$e^{+}e^{-}$ collider at DESY, where a lower limit was set  on $m_t>23.3$~GeV
in 1984. With the construction of the $Sp\bar{p}S$ at CERN,  and its
start-up in 1981, a new
energy frontier was explored  with the UA1 and UA2 experiments
searching for the top quark in $W$-boson decays. Both collaborations
set new lower limits on $m_t$, almost ruling out the possibility of
$W \rightarrow t\bar{b}$ decay. With the start-up of Tevatron
RunI, CDF and D0 joined the hunt for the top quark in 1992. Because
the detectors in RunI were quite different -- CDF being strong on tracking,
while D0, not having a solenoid magnet in the tracking system, had a better calorimeter than CDF --  the strategy to search for
the top quark was also different at the two experiments: CDF focused on
$b$-jet identification to reduce background, D0 used topological
information. The first
lower limits from CDF in 1992 of $m_t>91$~GeV finally eliminated the
possiblity of $W \rightarrow t\bar{b}$ decay. With the data collected
up to 1993, D0 also was able to set lower limits, now at
$m_t>131$~GeV, with a search focusing on $t\bar{t}$
production. In 1994, CDF published first evidence for $t\bar{t}$ events, where a small excess of top-like
events was observed. Although  D0 observed  a few top-like events (at
te predicted level), the luminosity was deemed insufficient to establish the existence of a top
quark. Finally, in winter 1994/1995, the data samples were large enough to
reach a conclusive result. On February 24th, 1995, CDF and D0
simultaneously submitted papers to Physics Review
Letters~\cite{cdftopdiscovery, d0topdiscovery},  based on 50~pb$^{-1}$ at
D0, and 67~pb$^{-1}$ at CDF, that announced definitive observations of
$t\bar{t}$ production. On March 2nd 1995, the
discovery was announced publicly 
in a joint session held  at the Fermilab auditorium. 

With the discovery of the top quark a large program was initiated to
study the particle's characteristics  in detail -- from its
production mechanism to
its inherent properties, as well as its connection to possible physics beyond
the standard model. An 
impressive understanding of the top quark has been achieved over the
past 16 years, some of which
will be summarized in the following section.

\section{The Present}
While the top quark discovery took place with just a handful of
events, the amount of data available today corresponds to thousands of
analyzable 
$t\bar{t}$ events per experiment, thereby enabling a precise measurement of
the $t\bar{t}$ production cross section, a detailed study of top quark
properties, and the performance of sensitive searches for new physics in the top quark
sector. Furthermore, the observation of electroweak single top-quark production in
2009~\cite{cdfsingletopdiscovery, d0singletopdiscovery}, also simultaneously by CDF and D0, provided another milestone in
the understanding of the top quark and the SM.
Almost everything we know  about the top quark has been 
achieved through pioneering studies performed at the Tevatron. These
relied not only on
the large amount of data, but also on the development and refinement of
 new ideas and analysis techniques. 
For example, the in-situ calibration of the JES using the hadronically
decaying $W$-boson and the Matrix Element (ME) method first realized at D0, has yielded
 the most precise measurements of the mass of the top quark, the
establishment of multivariate analysis techniques was crucial  in the search for
single top quarks,  and measurements of  the $t\bar{t}$ production cross
section have reached uncertainties
comparable to the uncertainty on the corresponding theoretical
quantum-chromodynamics 
predictions. Furthermore, several analyses have reached a precision
where systematic uncertainties are comparable or even larger than their
statistical components, as, for example, in  measurements of the top-quark mass
and in studies of the $W$ helicity fractions in $t\bar{t}$ decays.

All in all, the measurements performed  show that the
particle discovered in 1995 agrees well with the top quark predicted
by the SM. The production and properties are in full  accord with their
predicted values, and  a vast field of searches for signs beyond the
SM in states involving top quarks indicate no behaviour beyond the
SM. Recently, a deviation of measurements from
prediction in $t\bar{t}$ events was reported in the $t\bar{t}$
forward-backward asymmetry, showing higher measured asymmetries than expected.
While the LHC top-quark factory now provides more data   than the
Tevatron, many D0 and CDF analyses are still competitive, and some are
 complementary -- one being  the $t\bar{t}$ forward-backward
asymmetry. In the next section, I discuss  these competitive and
complementary analyses. 
A more detailed overview of the current understanding of top-quark production and its properties is given in several review papers,
e. g., Refs.~\cite{reviewpapers}, as well as in the articles in  these
proceedings.

\section{The Future}
Despite that the Tevatron collider is switched off and
the LHC experiments are collecting lage amounts of data, the Tevatron
analyses are proceeding at full pace. In the field of top quark
physics, the analyzed data corresponds to  about half of the 
collected data of about 10.5~fb$^{-1}$ at both CDF and D0. To decide which analyses are still
competitive with the new top quark factory, it is important to
understand the differences between the two colliders. Because a proton
and an antiproton are made to collide at the Tevatron, the initial
$p\bar{p}$ states are  CP-eigenstates, which is not true for two
protons undergoing collisions at the LHC. Furthermore, the energies in
the center of mass are different, with 
1.96~TeV at the Tevatron and currently 7~TeV at the LHC. At these
energies and respective types of interactions, the fraction of
$t\bar{t}$ events that are produced via $q\bar{q}$ annihilation and
via gluon-gluon fusion is about inverse: the former process happens 
about 85\% at the Tevatron, while it only contributes about 15\% at
the LHC. On the basis of these differences we can define  the legacy
measurements that appear to be important to pursue with the full Tevatron data. These comprise, in particular, the measurement of the production
cross sections -- differential and inclusive -- that will not be repeated at the same energy and
incident states, the measurement of $t\bar{t}$ spin correlations, which
are different in $q\bar{q}$ annihilation and gluon-gluon fusion, as
well as the $t\bar{t}$ forward-backward asymmetry, which does not
appear in the process $gg \rightarrow t\bar{t}$. Besides these
complementary studies, the measurement of the mass of the top quark
forms one of the legacy measurements of the 
Tevatron, as it benefits from a well
understood environment and is limited by systematic uncertainty. In
the following, I offer a brief perspective on each of these analyses.

\subsection{Complementarity resulting from the difference in energy and
  type of collisions: Production kinematics}
Although the $t\bar{t}$ production cross section at the LHC is a
factor of twenty
larger than at the Tevatron, a precise measurement of the $t\bar{t}$
production cross section as well as production kinematics, are
important to check the reliability of perturbative calculations in
quantum chromodynamics (QCD) as well as to search for
possible deviations from prediction that could suggest contributions
from  beyond the SM (BSM). While differential distributions  have thus
far been
measured only  with small statistics and for  few variables, the
extracted inclusive $t\bar{t}$ cross sections  for  5.6~fb$^{-1}$ of
data are highly dominated by systematic
uncertainties. It is likely therefore that measurements of the inclusive $t\bar{t}$ cross section
 will not be repeated in all final states.
The issue of  electroweak production of  single top quarks is quite
different, in that single top production can occur via
s-channel, t-channel or the Wt-channel. The latter has a negligible
cross section at the Tevatron, and will therefore not be measured
separately. The initial observation of single top quarks was realized
through the sum of s- and t-channel
contributions. But since BSM could affect the contributions differently in the different production modes, it is
important to disentangle the individual channels.
D0 recently reported the separate observation of the  t-channel
process~\cite{tchannelobservation}. The production in this channel has
also been reported recently by the LHC experiments ATLAS and
CMS~\cite{atlastchannel, cmstchannel}, where the production cross
section is  a factor of about 29  higher. For s-channel single-top
production, the enhancement factor at the LHC is only about 4.5, and
there  is far more background. The relatively
small s-channel cross section at the LHC and the fact that it
has not yet been observed separately, suggests this measurement for
priority at the
Tevatron.

\subsection{Complementarity resulting from the difference in the type of collisions: $t\bar{t}$ Spin
  Correlations}
By studying $t\bar{t}$ spin correlations, the full chain from
production to decay can be probed for consistency with the
SM. Predictions for  $t\bar{t}$ spin correlations depend on
the production mechanism, and on whether the top quarks are
produced at threshold. At the Tevatron, the main contribution
is from $q\bar{q}$ annihilation, where the spin of the top and
antitop quarks are parallel in the beam basis, while at the LHC,
events especially with
low $m_{t\bar{t}}$ are expected to have spins of top and antitop
mostly in 
  like-helicity configurations. The
  $t\bar{t}$ spin correlation at the Tevatron is therefore
  complementary to that at the LHC. Both CDF and D0 have explored
  $t\bar{t}$ spin correlations using both template and Matrix- Element techniques. Recently, first evidence for
  non-vanishing $t\bar{t}$ spin correlations was reported by D0 in
  5.4~fb$^{-1}$ of data in dilepton and lepton+jets final
  states~\cite{d0spinevidence}. As this measurement is still limited by
   statistics, the full data sample should  improve the precision of
   the result by a
  factor of $\sqrt{2}$, not counting  expected  improvements in the
 analyses.

\subsection{Complementarity resulting from the difference in the type of collisions: $t\bar{t}$
  Forward Backward Asymmetry}
One of the most interesting results this year is the 
forward-backward asymmetry of top and antitop production, where
measured asymmetries are consistently
larger than predicted from effects of QCD color charge. A particularly striking feature is the
enhanced asymmetry at large  $m_{t\bar{t}}$ reported by the CDF
collaboration. Besides trying to accommodate this apparent discrepancy
with theory, it is important to repeat the
statistically limited 
measurements using all available data. Furthermore, this kind of
effect is harder to measure at the LHC, as  it is easy to define the directions of the incoming quark and antiquark by
just assuming the proton and antiproton flight directions,
respectively, but  the
asymmetry also appears mainly in $q\bar{q}$ annihilation -- with far smaller
contributions from $qg$ states -- and not at all in gluon-gluon
fusion, which dominates at the LHC. Measurements at LHC are so far
based on  the difference in absolute rapidities of top and antitop
quarks, reflecting the widths of the two symmetric distributions, and
these effects are far more subtle and smaller than the predictions at
the Tevatron. By using the full data sample, the uncertainty 
on the measurement can be reduced by a factor of $\sqrt{2}$,
again ignoring  additional improvements expected  in the analysis. The
observed, unexpected behaviour puts the study of the $t\bar{t}$
forward-backward asymmetry as a high priority for continued  top-quark
analyses at the
Tevatron.

\subsection{Competitive because of a well understood environment: Mass
  of the top quark}
The mass of the top quark, $m_t$, is a free parameter in the SM, and, together with
the mass of the $W$ boson,  constrains  the mass of a SM Higgs boson. It
is therefore important to measure $m_t$ precisely. A variety of  methods have been invented at the
Tevatron to provide  the most precise mass measurements, as, for example template methods using neutrino or matrix
weighting, or the ME method. By performing  measurements  in all possible
final states,   using the wealth of techniques, and combining the
different results from CDF and D0, the uncertainty on $m_t$ has
reached $<1$~GeV for the first time this year~\cite{topmassaverage},
yielding $m_t=173.2\pm0.6 {\rm (stat)} \pm 0.8 {\rm (syst)}$~GeV. As this
result is limited by systematic uncertainties, the main task for the
experiments at LHC and Tevatron is to improve the understanding of systematic
contributions,  in particular, through studies of differences  in jet energy scale (JES)
for quark and gluon jets, as well as heavy-flavors,  and initial and final-state
radiation. Including some of the expected improvements on systematic
uncertainties, especially those related to JES, can reduce the
uncertainty on the combined Tevatron $m_t$  
to about 0.6~GeV in the final mass measurement from the Tevatron.

\section{Summary}
The Tevatron has provided many  technological
innovations and analysis techniques that have led to a huge
advancement in the understanding of fundamental particle physics, all of which define the legacy of the Tevatron. A
crucial part of the legacy is the discovery and detailed
characterization 
of the heaviest known elementary particle, the top quark, and the understanding of its production and properties 
achieved in the past 16 years of top quark physics at the Tevatron. The future of the top-quark program at the
Tevatron requires a priorization of measurements that are
complementary and competitive with the LHC. Besides the effort
to improve the analyses using all the collected data, the results from
each of the two
Tevatron experiments must be combined to provide the legacy of the
Tevatron.

\acknowledgments
I thank my collaborators from CDF and D0
 for their help in preparing the presentation and this
article, in particular Paul Grannis for providing material on the
history of the Tevatron and the top quark. I also thank the staffs at Fermilab and
collaborating institutions, and acknowledge the support from STFC.


\begin{thebibliography}{0}
\bibitem{TevHistory1}
H.~T.~Edwards,
  Ann.\ Rev.\ Nucl.\ Part.\ Sci.\ \ {\bf 35} (1985) 605.
\bibitem{TevHistory2}
 S.~Holmes, R.~S.~Moore and V.~Shiltsev,
  JINST\ {\bf 6} (2011) T08001.
\bibitem{TopHistory} \BY{B. Carithers and P. Grannis}
 http://www.slac.stanford.edu/pubs/beamline/pdf/95iii.pdf
\bibitem{cdftopdiscovery}   F.~Abe {\it et al.}  [CDF Collaboration],
  Phys.\ Rev.\ Lett.\  {\bf 74}, 2626 (1995).
\bibitem{d0topdiscovery}   S.~Abachi {\it et al.}  [D0 Collaboration],
  Phys.\ Rev.\ Lett.\  {\bf 74}, 2632 (1995).
\bibitem{cdfsingletopdiscovery} T. Aaltonen {\it et al.} [CDF
  Collaboration], Phys. Rev. Lett. {\bf 103}, 092002 (2009).
\bibitem{d0singletopdiscovery} V. M. Abazov {\it et al.} [D0
  Collaboration], Phys. Rev. Lett. {\bf 103}, 092001 (2009).
\bibitem{reviewpapers} A. B. Galtieri, F. Margaroli, I. Volobouev,
  arXiv:1109.2163;
D. Wicke, European Physical Journal C {\bf 71},
  1627 (2011); A. Heinson, Modern Physics Letters A {\bf 25}, 309
  (2010);  W.~Wagner,
 Mod.\ Phys.\ Lett.\ A\ {\bf 25}, 1297  (2010);  F. Fiedler {\it et al.}, Nuclear Instrum. Methods in
  Phys. Res. Sect. A 624, 203 (2010); F. Deliot and D. Glenzinski,
  arXiv:1010.1202v2; J. Incandela {\it et al.}, Progress in Particle
  and Nuclear Physics {\bf 63}, 239 (2009).
\bibitem{tchannelobservation} V. M. Abazov {\it et al.} [D0
  Collaboration],  Phys. Lett. B 705, 313 (2011).
\bibitem{atlastchannel} The ATLAS Collaboration, ATLAS-CONF-2011-101.
\bibitem{cmstchannel} The CMS Collaboration, 	Phys. Rev. Lett. {\bf
    107},  091802 (2011).
\bibitem{d0spinevidence} V. M. Abazov {\it et al.} [D0 Collaboration],   arXiv:1110.4194.
\bibitem{topmassaverage} The CDF and D0 Collaborations, arXiv:1107.5255.

\end{thebibliography}
\end{document}